\documentclass{./styles/svproc}
\usepackage{url}

\usepackage{textcomp,multirow}
\usepackage{soul}
\usepackage{color}

\usepackage{graphicx}
\usepackage[labelfont=bf]{caption}
\usepackage{amsmath}
\usepackage[table,x11names]{xcolor}
\usepackage[algo2e]{algorithm2e}
\usepackage{algorithmic}
\usepackage{algorithm}
\usepackage{placeins}
\usepackage{lipsum}
\usepackage{subcaption}
\usepackage[inline]{enumitem}
\usepackage{changepage}

\begin{document}
\mainmatter              
%

\title{Overview of Human Activity Recognition Using Sensor Data}
%
%

\author{Rebeen Ali Hamad \inst{1} \and Wai Lok Woo\inst{1},
Bo Wei\inst{2} \and Longzhi Yang\inst{1}}
%
%
%
\institute{Northumbria University, Newcastle Upon Tyne, NE1 8ST,UK,\\
\email{rebeen.hamad,wailok.woo,longzhi.yang@northumbria.ac.uk},
\and
Lancaster University, Lancaster,  LA1 4WA, UK, bo.wei@lancaster.ac.uk}

\maketitle              

\begin{abstract}

Human activity recognition (HAR) is an essential research field that has been used in different applications including home and workplace automation, security and surveillance as well as healthcare. Starting from conventional machine learning methods to the recently developing deep learning techniques and the Internet of things,  significant contributions have been shown in the HAR area in the last decade. Even though several review and survey studies have been published,  there is a lack of sensor-based HAR overview studies focusing on summarising the usage of wearable sensors and smart home sensors data as well as applications of HAR and deep learning techniques. Hence, we overview sensor-based HAR, discuss several important applications that rely on HAR, and highlight the most common machine learning methods that have been used for HAR. Finally, several challenges of HAR are explored that should be addressed to further improve the robustness of HAR.

\end{abstract}
\section{Introduction}
\begin{figure}[!t]
\includegraphics[width=\textwidth]{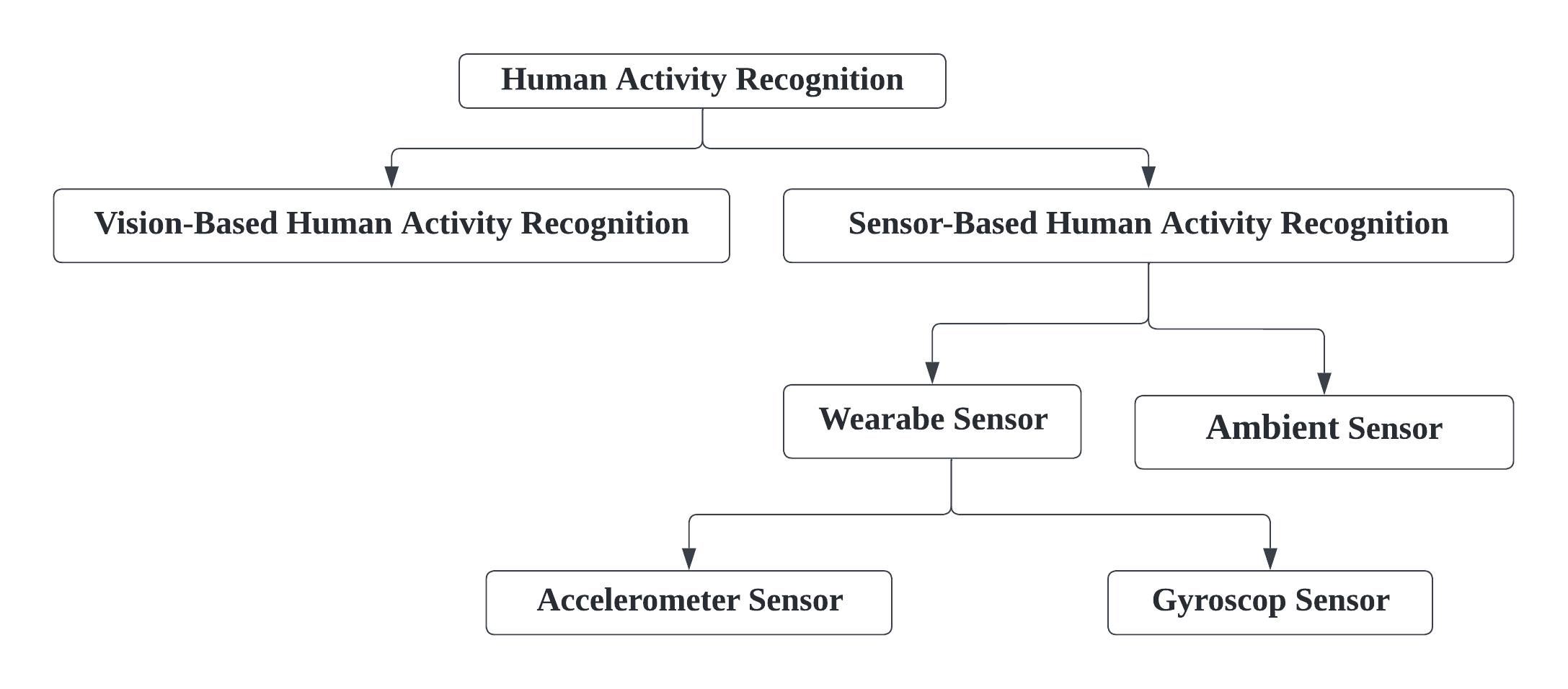}
\caption{ Taxonomy of Human Activity Recognition } \label{fig:harbased}
\end{figure}

Sensor-based HAR refers to automatically recognizing human activities from collected data generated by different sensing devices including wearable and ambient sensors (smart home environment sensors).  HAR based on sensors data has been utilized in different fields of study such as healthcare systems, behaviour analysis, ambient assisted living (AAL), patient monitoring systems \cite{qi2018hybrid,aviles2019granger,sankar2018internet,capela2015feature}.  HAR is often classified into two main categories as shown in Figure \ref{fig:harbased}: sensor-based recognition and vision-based recognition.  Vision-based HAR methods utilize one or several cameras for recording video examples of human activities.  Moreover, multiple views of visual human activities are used to detect human movements.  However,  people are generally reluctant to use cameras for recording daily activity data due to privacy concerns  \cite{jung2020review}.  Another limitation is that processing visual data for HAR using cameras could be computationally expensive. Unlike vision-based HAR, sensor-based HAR has gained more outstanding acceptability in the users and research communities due to low cost and privacy protection \cite{jung2020review}.  Besides,  rapid evolutions in sensor technologies and ubiquitous computing have enabled sensors-based HAR with a satisfactory performance at a lower computational cost \cite{jung2020review}.  

HAR using machine learning methods from a series of data captured by sensors provides insights into what people are performing such as walking, showering, eating,  or sleeping \cite{hamad2020efficacy}.  Conventional machine learning methods have shown great progress and reached satisfying performance on HAR. Such methods include Na\"ive Bayes (NB), Support Vector Machine (SVM), Hidden Markov Model (HMM),  k-Nearest Neighbour (KNN),  Decision Tree (DT), and Random Forest (RF) \cite{anjum2013activity}.  However, conventional methods rely heavily on handcrafted heuristic feature extraction, which is mostly domain-dependent, expensive, and usually needs domain experts \cite{hamad2020efficacy}. Moreover,  handcrafted features are mostly specific to a  domain and often less generalizable for application domains. Handcrafted features cannot develop an adequate amount of features from raw sensors data and are time-consuming  \cite{hamad2020efficacy}. Due to the aforementioned issues, conventional machine learning algorithms have become less popular. Thereby, the power of automated feature extraction methods has received increasingly remarkable attention. Hence, more effective, capable and efficient deep learning algorithms have been developed for HAR systems \cite{hamad2019efficient}. The most popular methods of deep learning include Recurrent Neural Networks (RNN),  Long Short-Term Memory (LSTM) and Convolution Neural Networks (CNN) \cite{hamad2019efficient}. The sensors deployed in HAR systems could be broadly categorized into two modalities: wearable sensor-based HAR  and ambient sensors-based HAR \cite{hamad2019efficient} as shown in Figure \ref{fig:harbased}.

\section{HAR Based on Wearable Sensor Data}

Wearable sensors are unobtrusive, portable and inexpensive devices which can be conveniently worn by individuals. Wearable sensors are designed to meet some typical essentials including low battery consumption, small size, and high measurement accuracy. Furthermore, wearable sensors have recently received considerable attention and gained the potential to provide assisted living in healthcare applications and particularly to monitor human activities \cite{cicirelli2021ambient}. Wearable sensors are one of the most prevalent modalities in HAR systems.  The wearable sensors generate a continuous stream of information based on the changes in the acceleration and angular velocity of human body movements. Thus the generated informative data are used to recognise human activities using machine learning methods \cite{cicirelli2021ambient}. The wearable sensors can be embedded in a human body and distributed from head to foot.  The wearable sensors can work in a moderately wide area while the wearer is performing activities such as walking and running \cite{cicirelli2021ambient}.   Wearable sensors such as accelerometers and gyroscopes are often embedded in smartphones, smartwatches,  smart bands, clothes, belts, glasses, helmets, or shoes \cite{cicirelli2021ambient}.  Table \ref{table:wearablesensor} shows a review of HAR based on wearable sensors.

\subsection{Accelerometer Wearable Sensors}

Wearable sensors mainly consist of accelerometers and gyroscopes that have been broadly used to capture and extract diverse information related to the human motion for HAR \cite{hamad2021dilated}. 
Accelerometers have been deployed for different applications such as fall detection \cite{shany2011sensors}, body motion analysis and movement \cite{kan2011wearable},  individual's postural orientation \cite{sazonov2010monitoring}, and assessment of people with Parkinson's disease  \cite{mariani2012shoe}. Human daily activities such as standing, cycling, walking, running, sitting, and walking downstairs and upstairs using accelerometer data can be effectively recognized. In 2004,  Bao and Intille \cite{bao2004activity} employ numerous accelerometers to collect data from 20 subjects and to recognise 20 activities such as walking, brushing teeth,  Reading, climbing stairs and vacuuming in a realistic setting. Since then, considerable systems of human activity recognition based on accelerometer sensor data have followed \cite{cornacchia2016survey}. Accelerometers are the most commonly employed body-worn sensors in HAR systems due to their ability to render data based on the motion of the wearer   \cite{cornacchia2016survey}. 



Smartphones as wearable sensors have been advancing rapidly and attracting great attention for HAR due to having various built-in sensing units including gyroscopes, accelerometers,  cameras, and Global Positioning System (GPS) sensors \cite{ronao2016human}.  Smartphones can be easily exploited to collect data instead of using different wearable sensors since Smartphones could be easily placed on different parts of an individual's body ranging from the upper such as the arm or wrist to the leg or ankle the lower \cite{chen2019smartphone}.  Moreover, smartphones can be utilized in both indoor and outdoor settings to record daily physical activities for  HAR systems.  The aforementioned features of smartphones for collecting human daily activities have increased the capability of the HAR system.  The accelerometer sensors data are used for HAR systems more than other sensors data from smartphone sensors \cite{chen2012sensor}. Furthermore,  the features of smartphones over other wearable devices have made smartphones a key ubiquitous platform for HAR systems \cite{chen2012sensor}. The features are first: a smartphone is a low-cost machine that provides different software and hardware sensors in a single device. Second, smartphones can capture and process data since they are programmable devices. Moreover, smartphones are able to transmit and receive data as well as connect with other devices that have made smartphones an ideal platform for HAR systems in the research community \cite{sousa2019human}.

\renewcommand{\arraystretch}{1.}
\begin{table}[h!]
\setlength{\tabcolsep}{6.4pt}
 \begin{adjustwidth}{-1.1cm}{}
\begin{center}
\caption{Review of HAR based on wearable sensors } \label{table:wearablesensor}

 \begin{tabular}{|m{9em}|l| m{9em}|m{18em}|} 
 \hline
 References &
\rotatebox[origin=c]{90}{{ No.Sensors}} 
   &  Sensor placements &  Activities  \\

 \hline 
Zheng  et al. \cite{zheng2013physical}  & 1 & Waist, Wrist, Hip
Pocket &   Standing, Lying, Walking, Dancing, Running, Upstairs, Downstairs, Jogging, Skipping, Sitting  \\
 \hline 
  
Gjoreski and Gams \cite{gjoreski2011accelerometer}&7& Right ankle, Left thigh, Chest &  Sitting, Going down,   Standing up, Lying, Sitting on the ground, Standing\\
\hline 
Jiang et al. \cite{jiang2011method}   &4  & Right forearm, Right shank, Left forearm, Left shank &  Standing, Walking on an elliptical machine, Sitting, Running on an elliptical machine, Lying on a bed,  Cycling, Jogging,  Rowing, Walking and weight lifting \\
\hline 
Jennifer et al. \cite{kwapisz2011activity} & 1&Smartphone  & Sitting, Jogging, Upstairs, Walking, Downstairs, Standing   \\
\hline 
 Chun and Weihua \cite{zhu2011motion}   & 1&  Right thigh &Stand-to-sit, Sitting, Lie-to-sit, Standing, Lying, Sit-to-stand, Walking, Sit-to-lie   \\
\hline 
Siirtola and Roning, 2012 \cite{siirtola2012user} & 1&Smartphone
placed in pocket & Cycling, Walking, Driving a car ,  Standing, Sitting  \\

\hline
Sweetlin  \cite{hemalatha2013frequent} & 1& Chest & Fall, Walking, Standing, Lying , Sitting   \\
  \hline
  
 Mannini et al. \cite{mannini2013activity}   & 1& Wrist, Ankle   & 26 daily activities  \\
 \hline

Lei et al.\cite{gao2014evaluation}& 4 &Chest, Left under-arm, Waist and thigh & Standing, Stand-to-sit, walking,Sitting, Lie-to-stand, Lying, Sit-to-stand, Downstairs Stand-to-lie\\
\hline

Ronao, C.A. and Cho, S.B. \cite{ronao2016human}&1&Kept in the pocket& Walking, Laying, Sitting, Walking upstairs, Down upstairs, Standing\\

\hline
Davila JC et al. \cite{davila2017wearable}&19& Hands, Left Foot, Back, upright knee, right foot, low right knee, Hip,  & Standing, Sitting, Walking, Lie  \\
\hline

Hassan MM et al. \cite{hassan2018robust}& 1& Smartphone: kept in the pocket& Standing, Stand-to-Lie, Walking,  Stand-to-Sit, Lying down, Lie-to-Sit, Sitting, Lie-to-Stand, Walking-upstairs, Sit-to-Stand,  Walking-downstairs, Sit-to-Lie 
\\
\hline

Wan S et al. \cite{wan2020deep}&1&Smartphone: kept in the pocket&
 Housecleaning, Sitting, Folding,  Running,  Laundry,  Lying, Playing soccer, Walking, Computer work, Standing\\
\hline

Mekruksavanich S et al. \cite{mekruksavanich2021lstm}&1& Smartphone: kept in the pocket& Sitting, Walking downstairs, Laying, Walking upstairs, Standing\\

\hline

Han C et al. \cite{han2022human} &1&Smartphone: kept in the pocket&
Going upstairs, Going downstairs, Jogging, Jumping and Walking\\

\hline
\end{tabular}
\end{center}
\end{adjustwidth}
\vspace*{-5mm}
\end{table} 

\subsection{Gyroscope Wearable Sensors}

Gyroscope sensors can be used to measure angular velocity and preserve the orientation of an object.  The difference in the angle could be detected over a period of time by comparing the initial know value with the change of the angle. The limitations of gyroscope sensors are output drift over time and the sensitivity of the gyroscope to a certain range of angular velocities. Often accelerometer sensors have been used in human activity recognition or the combination of gyroscope and accelerometer sensors. Narayanan et al.\cite{narayanan2009longitudinal} performed estimating the fall risk of the movements of  68 older adults to measure the timed sit-to-stand, up-and-go test. Besides five more repetitions and alternate step tests are conducted using a triaxial accelerometer attached to the waist. Greene et al. \cite{greene2010quantitative}  used both gyroscopes and accelerometers attached to two legs of a user to measure the timed up-and-go test to differentiate non-fallers from fallers. Varkey et al.\cite{varkey2012human} used both gyroscopes and accelerometers and attached these sensors to the right foot and right arm wrist of the user to obtain linear and angular accelerations. Regarding activity monitoring, inertial sensors can be useful due to these features low power requirements, low cost,  compact size, non-intrusiveness, and the ability to provide data directly associated with the movement of the user. However, inertial sensors have these limitations. Firstly the placement of the inertial sensor on various parts of the human body causes an uncomfortable feeling for older adults which may lead to low acceptance by the people.  Secondly, due to collecting data continuously by the inertial sensors, battery life could be reduced.  inertial sensors cannot provide adequate information for monitoring complex movements and activities that involve many human-object interactions \cite{varkey2012human}.

\section{HAR Based on Smart Homes Sensor Data}

The steady increase in elderly people's population has been identified as a prominent social problem and financial challenge for the future decade including increased elderly healthcare demands, financial stress and unbalanced supply-demand.  Consequently, the demand for caregivers to provide care for elderly people, expenditure on healthcare costs,  and the desire of older adults to live unassisted in their residences have increased \cite{hamad2022cross}. Smart home environments provide one of the most promising solutions to deliver ubiquitous and context-aware services and monitor activities of daily life.   Smart homes are equipped with diverse types of sensors to help older individuals who wish to remain in their residences.  Various sensors have been deployed in smart homes to capture various human activities. For instance, passive infrared (PIR) sensors to monitor the motions of the resident, reed switches on cupboards and doors to detect open or close status, and pressure sensors on beds, chairs and sofas to detect the presence of residents.  Moreover, float sensors in the washroom to detect whether the flush is performed or not \cite{hamad2022cross}. Remarkably,  with the rapid development of new low-cost sensors,  cloud computing and the emerging  Internet of Things (IoT) technologies, AAL systems deployed
in smart homes have intelligently been used to monitor, record and act on physical surroundings and potentially support different applications such as health and wellness evaluation, short and long-term activity pattern analyses, consistent rehabilitation instruction, fall detection at home,  chronic disease management, and timely medication reminder  \cite{hamad2019efficient}. For senior people who hope to remain independently and functionally in their own home, they should have enough ability to perform Activities of Daily Living (ADL) such as preparing breakfast and lunch, drinking, cooking,  eating, and washing grooming to be used by the caregivers to know a
resident's functional status \cite{hamad2019efficient}.  Smart home environments have been employed to unobtrusively observe the interactions of a user with the physical surrounding objects in the environments and the user's activities \cite{hamad2019efficient}. The interactions are the ADL and conducted by an individual in a particular place with the specific object in the environment which have been processed using machine learning methods to detect activities \cite{hamad2022cross}. For example,  cooking activity takes place in the kitchen,  preparing dinner or breakfast are usually takes place in the kitchen, shower related activities such as brushing teeth takes place in the bathroom. Thus, environmental sensors are deployed to capture an individual's existence and interaction with the physical surrounding objects and furniture (e.g.,coffee table, dining table, chairs, cupboard, bed, sofa) \cite{storf2009event}. For instance, if the sensor readings indicate a stove, toaster, or microwave is working, and the door of the refrigerator or the cupboard is opened, then an activity is taking place in the kitchen.  Hence based on the objects or furnitures usage the activities will be distinguished and detected.  Thus,  correctly recognizing in-home human daily activities plays a significant role in comprehending and analysing the association between residents and their physical surroundings to help residents stay safe and healthy in their own independently residence and reduce the costs of healthcare \cite{wang2020activities}.

\subsection{ Smart Environment Embedded Sensors}

The most typical sensors that have been deployed in smart homes to record human activities will be described. Binary sensors are often used in smart environments that generate 0 or 1 to capture changes such as "door opened" or "door closed". Binary sensors such as pressure sensors, motion sensors,  contact switch sensors, and state-change sensors have been deployed on a different object to monitor and capture interactions of a resident with the physical surroundings in a particular place of the environment \cite{hamad2018stability}.

\begin{itemize}
\item Motion sensors such as Passive InfraRed (PIR) \cite{vera2016real} have been being widely used to track and record the presence of a resident in a  particular place of the smart environment, for example, bed room, kitchen, or bathroom. 

\item A simple state-change sensor deployed to detect human object interactions by recording the  changes of the state of the objects 
 \cite{ordonez2013online}.  For instance,  state-change sensor can  be bound to the handset of a home telephone to capture the interactions between a resident and the home telephone when the resident lifts the handset from the telephone base station in the smart environment.
 
\item  Pressure sensors are utilised to detect sitting or lying on chairs or beds to unobtrusively monitor and record the presence and absence of a resident \cite{chen2012sensor}.

\item Contact switch sensors can be connected to the different objects including the doors of beds, living or kitchen rooms. Further,  this type of sensor has been linked with the fridge door, or cabinets to detect interactions between residents with the physical surrounding objects \cite{ding2011sensor}. 
\end{itemize}

\section{Applications of Human Activity Recognition}
In this section,  applications that can rely on HAR are highlighted. Particularly, we review healthcare,  security and surveillance,  entertainment and games, and home automation applications.  For each of these applications, we present the benefits of automatically recognizing human activities that can deliver useful services. 

\subsection{Healthcare Applications}
HAR  plays a crucial role in assisting the physical and mental well-being of the population. HAR can be used to conduct robust recognition of unsafe situations and detect deviations of behaviour to improve older adults' care are alert systems \cite{hamad2020joint}. Furthermore, HAR could be potentially used to reduce or prevent the risk of various chronic diseases such as diabetes, obesity, neurological conditions, and cardiovascular \cite{ogbuabor2018human}. People with these diseases in addition to their treatment are usually following an effective physical activity scheme or routines such as cycling, walking, running jogging. Accurate HAR can help caregivers to identify whether the patient or observee has any difficulties to follow the routines to perform activities.  Besides, adequate information regarding the duration of activities is useful for individuals to compliance their ADL according to prescription and for practitioners to monitor and assess the health status. HAR in a healthcare application is a proactive process to adopt  healthy lifestyle for people who follow a daily routines, for example everyday activities including exercising, sleeping, and social relationships \cite{chelli2019machine}.
\subsection{Security and Surveillance}

HAR has been broadly employed for surveillance systems based on vision and sensor data. Multiple solutions for HAR-based visual data have been proposed to detect various suspicious activities in a public place \cite{saba2021suspicious}. Besides sensor data streams are also used for security purposes.  Sun et al. \cite{sun2017sos} proposed a HAR surveillance system to detect various physical assaults and criminal offences such as gunshots, abuse, and kidnapping.

\subsection{Home and workplace Automation}

HAR systems have been employed to control appliances within the home and offices.  HAR used to turn on devices in a room when a resident entered the room automatically and also the devices are turned off when the resident left the room to save energy and increase safety \cite{antar2019challenges}.

\section{Deep learning for HAR}

Conventional machine learning methods such as SVM, NB, HMM, RF, and KNN have made remarkable progress on HAR and rendered reasonable outcomes \cite{sousa2019human}. 
However, due to the intrinsic complexity of physical human activities, traditional machine learning models may not successfully learn non-linear relationships among sensor generated data.  Besides, the  traditional machine learning algorithms rely on the handcraft features which needs domain knowledge experts which is mostly expensive and time consuming. Furthermore, feature extraction become a pre step for classification thus leads to sub-optimization. In contrast to traditional machine learning methods, deep learning models are able to  automatically learn complex features from the sensors generated data and  jointly optimize both feature extraction  and classifier learning \cite{sousa2019human}.Among the deep learning methods, LSTM  and one-dimensional CNN as sequential deep learning models have been commonly proposed for HAR  based on smart home sensors and wearable sensors data. LSTM  has obtained  state-of-the-art performance for temporal information processing in various applications particularly HAR based on sensors data. LSTM process data using gating mechanism that allows long-term dependencies. 1D CNN for human activity recognition has particularly two advantages which are local dependency and scale invariance. Local dependency referes to  local movement patterns withing an activity form the sequence input data. Besides, scale invariance referes to the CNN’s power in  detecting  activity patterns even when the activity motion changes in some way, for instance, a individual may  run with various movement intensity \cite{hamad2019efficient}.

\section{Challenges of HAR}
In this section, several challenges of HAR have been explored that stem from the reviewed literature.

\begin{enumerate}[label=\roman*.]
  \item Annotation Human Activities: one of the limitations of employing the aforementioned machine learning algorithms is to annotate an adequately amount of data since this procedure requires domain knowledge experts as well as is time-consuming and expensive. Moreover, particularly, labelling sensor data for HAR compared to labelling image data needs extra steps by recording human activities with a synchronized video to correctly annotate raw sensor human activities \cite{chen2021deep,hamad2022cross}. 
  \item Accurate HAR: several human activities are difficult to correctly detect because activities are inherently composed of several small actions and these actions have a different order or duration for different individuals. Mostly HAR  has been employed to detect what activity is performed at a specific time rather than how the activity is performed \cite{ni2015elderly,dang2020sensor}.
  
  \item  Imbalanced Class Problem: performing some activities occur more frequently, particularly for elderly individuals with diabetes such as sleeping, drinking and eating compared to walking, or running. This generates imbalanced datasets where some classes (minority class) have a very low number of samples compared to other classes (majority class).  The imbalanced class problem often occurs when multiple activities are detected using multiple sensors \cite{ni2015elderly,antar2019challenges}.  
  
  \item Overlapping Activities: several human activities are similar and share some characteristics such as snack, breakfast, lunch or dinner that cause the overlapping problem. These activities are less discriminative due to their overlaps in the feature space which makes the recognition process much harder. 
  
  \item Multi-Sensor and Multi-user Activity Recognition: processing various sensor data to generate adequate accuracy is still an open research problem.  Besides, often HAR is performed for single-user activities, however, in real-life scenarios, concurrently multiple activities could be conducted by numerous individuals. Recognizing multi-user activities and their interactions are clearly more challenging \cite{ni2015elderly}.  
  
 \end{enumerate}

\section{Conclusion}
This paper presents an overview of the HAR area, focusing on wearable and smart home sensors. This paper firstly discussed the concept of HAR followed by an introduction of the HAR  based on wearable sensors and smart home data.  The introduction described the different wearable sensors and smart home sensors that have been used for HAR systems.
 The aim of the introduction is to support the scientific community in the-state-of-the-art for the HAR area in the context of wearable sensors and smart home sensors.  Besides,  this paper discussed the applications that rely on activity recognition and highlighted the most common machine learning algorithms that have been used for HAR. Finally, this paper identified some of the significant research challenges of HAR that should be addressed to further improve the performance, robustness and precision of HAR. 

\bibliographystyle{unsrt}

\begin{thebibliography}{10}

\bibitem{qi2018hybrid}
Jun Qi, Po~Yang, Martin Hanneghan, Stephen Tang, and Bo~Zhou.
\newblock A hybrid hierarchical framework for gym physical activity recognition
  and measurement using wearable sensors.
\newblock {\em IEEE Internet of Things Journal}, 6(2):1384--1393, 2018.

\bibitem{aviles2019granger}
Carlos Aviles-Cruz, Eduardo Rodriguez-Martinez, Juan Villegas-Cortez, and
  Andr{\'e}s Ferreyra-Ramirez.
\newblock Granger-causality: An efficient single user movement recognition
  using a smartphone accelerometer sensor.
\newblock {\em Pattern Recognition Letters}, 125:576--583, 2019.

\bibitem{sankar2018internet}
S~Sankar, P~Srinivasan, and R~Saravanakumar.
\newblock Internet of things based ambient assisted living for elderly people
  health monitoring.
\newblock {\em Research Journal of Pharmacy and Technology}, 11(9):3900--3904,
  2018.

\bibitem{capela2015feature}
Nicole~A Capela, Edward~D Lemaire, and Natalie Baddour.
\newblock Feature selection for wearable smartphone-based human activity
  recognition with able bodied, elderly, and stroke patients.
\newblock {\em PloS one}, 10(4):e0124414, 2015.

\bibitem{jung2020review}
Im~Y Jung.
\newblock A review of privacy-preserving human and human activity recognition.
\newblock {\em International Journal on Smart Sensing \& Intelligent Systems},
  13(1), 2020.

\bibitem{hamad2020efficacy}
Rebeen~Ali Hamad, Masashi Kimura, and Jens Lundstr{\"o}m.
\newblock Efficacy of imbalanced data handling methods on deep learning for
  smart homes environments.
\newblock {\em SN Computer Science}, 1(4):1--10, 2020.

\bibitem{anjum2013activity}
Alvina Anjum and Muhammad~U Ilyas.
\newblock Activity recognition using smartphone sensors.
\newblock In {\em 2013 ieee 10th consumer communications and networking
  conference (ccnc)}, pages 914--919. IEEE, 2013.

\bibitem{hamad2019efficient}
Rebeen~Ali Hamad, Alberto~Salguero Hidalgo, Mohamed-Rafik Bouguelia,
  Macarena~Espinilla Estevez, and Javier~Medina Quero.
\newblock Efficient activity recognition in smart homes using delayed fuzzy
  temporal windows on binary sensors.
\newblock {\em IEEE journal of biomedical and health informatics},
  24(2):387--395, 2019.

\bibitem{cicirelli2021ambient}
Grazia Cicirelli, Roberto Marani, Antonio Petitti, Annalisa Milella, and
  Tiziana D’Orazio.
\newblock Ambient assisted living: A review of technologies, methodologies and
  future perspectives for healthy aging of population.
\newblock {\em Sensors}, 21(10):3549, 2021.

\bibitem{hamad2021dilated}
Rebeen~Ali Hamad, Masashi Kimura, Longzhi Yang, Wai~Lok Woo, and Bo~Wei.
\newblock Dilated causal convolution with multi-head self attention for sensor
  human activity recognition.
\newblock {\em Neural Computing and Applications}, 33(20):13705--13722, 2021.

\bibitem{shany2011sensors}
Tal Shany, Stephen~J Redmond, Michael~R Narayanan, and Nigel~H Lovell.
\newblock Sensors-based wearable systems for monitoring of human movement and
  falls.
\newblock {\em IEEE Sensors Journal}, 12(3):658--670, 2011.

\bibitem{kan2011wearable}
Yao-Chiang Kan and Chun-Kai Chen.
\newblock A wearable inertial sensor node for body motion analysis.
\newblock {\em IEEE Sensors Journal}, 12(3):651--657, 2011.

\bibitem{sazonov2010monitoring}
Edward~S Sazonov, George Fulk, James Hill, Yves Schutz, and Raymond Browning.
\newblock Monitoring of posture allocations and activities by a shoe-based
  wearable sensor.
\newblock {\em IEEE Transactions on Biomedical Engineering}, 58(4):983--990,
  2010.

\bibitem{mariani2012shoe}
Benoit Mariani, Mayt{\'e}~Castro Jim{\'e}nez, Fran{\c{c}}ois~JG Vingerhoets,
  and Kamiar Aminian.
\newblock On-shoe wearable sensors for gait and turning assessment of patients
  with parkinson's disease.
\newblock {\em IEEE transactions on biomedical engineering}, 60(1):155--158,
  2012.

\bibitem{bao2004activity}
Ling Bao and Stephen~S Intille.
\newblock Activity recognition from user-annotated acceleration data.
\newblock In {\em International conference on pervasive computing}, pages
  1--17. Springer, 2004.

\bibitem{cornacchia2016survey}
Maria Cornacchia, Koray Ozcan, Yu~Zheng, and Senem Velipasalar.
\newblock A survey on activity detection and classification using wearable
  sensors.
\newblock {\em IEEE Sensors Journal}, 17(2):386--403, 2016.

\bibitem{ronao2016human}
Charissa~Ann Ronao and Sung-Bae Cho.
\newblock Human activity recognition with smartphone sensors using deep
  learning neural networks.
\newblock {\em Expert systems with applications}, 59:235--244, 2016.

\bibitem{chen2019smartphone}
Zhenghua Chen, Chaoyang Jiang, Shili Xiang, Jie Ding, Min Wu, and Xiaoli Li.
\newblock Smartphone sensor-based human activity recognition using feature
  fusion and maximum full a posteriori.
\newblock {\em IEEE Transactions on Instrumentation and Measurement},
  69(7):3992--4001, 2019.

\bibitem{chen2012sensor}
Liming Chen, Jesse Hoey, Chris~D Nugent, Diane~J Cook, and Zhiwen Yu.
\newblock Sensor-based activity recognition.
\newblock {\em IEEE Transactions on Systems, Man, and Cybernetics, Part C
  (Applications and Reviews)}, 42(6):790--808, 2012.

\bibitem{sousa2019human}
Wesllen Sousa~Lima, Eduardo Souto, Khalil El-Khatib, Roozbeh Jalali, and Joao
  Gama.
\newblock Human activity recognition using inertial sensors in a smartphone: An
  overview.
\newblock {\em Sensors}, 19(14):3213, 2019.

\bibitem{zheng2013physical}
Yonglei Zheng, Weng-Keen Wong, Xinze Guan, and Stewart Trost.
\newblock Physical activity recognition from accelerometer data using a
  multi-scale ensemble method.
\newblock In {\em Twenty-Fifth IAAI Conference}, 2013.

\bibitem{gjoreski2011accelerometer}
Hristijan Gjoreski and Matja{\v{z}} Gams.
\newblock Accelerometer data preparation for activity recognition.
\newblock In {\em Proceedings of the International Multiconference Information
  Society, Ljubljana, Slovenia}, volume 1014, page 1014, 2011.

\bibitem{jiang2011method}
Ming Jiang, Hong Shang, Zhelong Wang, Hongyi Li, and Yuechao Wang.
\newblock A method to deal with installation errors of wearable accelerometers
  for human activity recognition.
\newblock {\em Physiological measurement}, 32(3):347, 2011.

\bibitem{kwapisz2011activity}
Jennifer~R Kwapisz, Gary~M Weiss, and Samuel~A Moore.
\newblock Activity recognition using cell phone accelerometers.
\newblock {\em ACM SigKDD Explorations Newsletter}, 12(2):74--82, 2011.

\bibitem{zhu2011motion}
Chun Zhu and Weihua Sheng.
\newblock Motion-and location-based online human daily activity recognition.
\newblock {\em Pervasive and Mobile Computing}, 7(2):256--269, 2011.

\bibitem{siirtola2012user}
Pekka Siirtola and Juha R{\"o}ning.
\newblock User-independent human activity recognition using a mobile phone:
  Offline recognition vs. real-time on device recognition.
\newblock In {\em Distributed computing and artificial intelligence}, pages
  617--627. Springer, 2012.

\bibitem{hemalatha2013frequent}
C~Sweetlin Hemalatha and Vijay Vaidehi.
\newblock Frequent bit pattern mining over tri-axial accelerometer data streams
  for recognizing human activities and detecting fall.
\newblock {\em Procedia Computer Science}, 19:56--63, 2013.

\bibitem{mannini2013activity}
Andrea Mannini, Stephen~S Intille, Mary Rosenberger, Angelo~M Sabatini, and
  William Haskell.
\newblock Activity recognition using a single accelerometer placed at the wrist
  or ankle.
\newblock {\em Medicine and science in sports and exercise}, 45(11):2193, 2013.

\bibitem{gao2014evaluation}
Lei Gao, AK~Bourke, and John Nelson.
\newblock Evaluation of accelerometer based multi-sensor versus single-sensor
  activity recognition systems.
\newblock {\em Medical engineering \& physics}, 36(6):779--785, 2014.

\bibitem{davila2017wearable}
Juan~Carlos Davila, Ana-Maria Cretu, and Marek Zaremba.
\newblock Wearable sensor data classification for human activity recognition
  based on an iterative learning framework.
\newblock {\em Sensors}, 17(6):1287, 2017.

\bibitem{hassan2018robust}
Mohammed~Mehedi Hassan, Md~Zia Uddin, Amr Mohamed, and Ahmad Almogren.
\newblock A robust human activity recognition system using smartphone sensors
  and deep learning.
\newblock {\em Future Generation Computer Systems}, 81:307--313, 2018.

\bibitem{wan2020deep}
Shaohua Wan, Lianyong Qi, Xiaolong Xu, Chao Tong, and Zonghua Gu.
\newblock Deep learning models for real-time human activity recognition with
  smartphones.
\newblock {\em Mobile Networks and Applications}, 25(2):743--755, 2020.

\bibitem{mekruksavanich2021lstm}
Sakorn Mekruksavanich and Anuchit Jitpattanakul.
\newblock Lstm networks using smartphone data for sensor-based human activity
  recognition in smart homes.
\newblock {\em Sensors}, 21(5):1636, 2021.

\bibitem{han2022human}
Chaolei Han, Lei Zhang, Yin Tang, Wenbo Huang, Fuhong Min, and Jun He.
\newblock Human activity recognition using wearable sensors by heterogeneous
  convolutional neural networks.
\newblock {\em Expert Systems with Applications}, 198:116764, 2022.

\bibitem{narayanan2009longitudinal}
Michael~R Narayanan, Stephen~J Redmond, Maria~Elena Scalzi, Stephen~R Lord,
  Branko~G Celler, Nigel~H Lovell, et~al.
\newblock Longitudinal falls-risk estimation using triaxial accelerometry.
\newblock {\em IEEE Transactions on Biomedical Engineering}, 57(3):534--541,
  2009.

\bibitem{greene2010quantitative}
Barry~R Greene, Alan O’Donovan, Roman Romero-Ortuno, Lisa Cogan, Cliodhna~Ni
  Scanaill, and Rose~A Kenny.
\newblock Quantitative falls risk assessment using the timed up and go test.
\newblock {\em IEEE Transactions on Biomedical Engineering}, 57(12):2918--2926,
  2010.

\bibitem{varkey2012human}
John~Paul Varkey, Dario Pompili, and Theodore~A Walls.
\newblock Human motion recognition using a wireless sensor-based wearable
  system.
\newblock {\em Personal and Ubiquitous Computing}, 16(7):897--910, 2012.

\bibitem{hamad2022cross}
Rebeen~Ali Hamad, Longzhi Yang, Wai~Lok Woo, and Bo~Wei.
\newblock Cross-domain activity recognition using shared representation in
  sensor data.
\newblock {\em IEEE Sensors Journal}, 2022.

\bibitem{storf2009event}
Holger Storf, Thomas Kleinberger, Martin Becker, Mario Schmitt, Frank Bomarius,
  and Stephan Prueckner.
\newblock An event-driven approach to activity recognition in ambient assisted
  living.
\newblock In {\em European conference on ambient intelligence}, pages 123--132.
  Springer, 2009.

\bibitem{wang2020activities}
Aiguo Wang, Shenghui Zhao, Chundi Zheng, Jing Yang, Guilin Chen, and Chih-Yung
  Chang.
\newblock Activities of daily living recognition with binary environment
  sensors using deep learning: A comparative study.
\newblock {\em IEEE Sensors Journal}, 21(4):5423--5433, 2020.

\bibitem{hamad2018stability}
Rebeen~Ali Hamad, Eric J{\"a}rpe, and Jens Lundstr{\"o}m.
\newblock Stability analysis of the t-sne algorithm for human activity pattern
  data.
\newblock In {\em 2018 IEEE international conference on systems, man, and
  cybernetics (SMC)}, pages 1839--1845. IEEE, 2018.

\bibitem{vera2016real}
Alejandro Vera-Baquero, Ricardo Colomo-Palacios, and Owen Molloy.
\newblock Real-time business activity monitoring and analysis of process
  performance on big-data domains.
\newblock {\em Telematics and Informatics}, 33(3):793--807, 2016.

\bibitem{ordonez2013online}
Fco~Javier Ord{\'o}{\~n}ez, Jos{\'e}~Antonio Iglesias, Paula De~Toledo, Agapito
  Ledezma, and Araceli Sanchis.
\newblock Online activity recognition using evolving classifiers.
\newblock {\em Expert Systems with Applications}, 40(4):1248--1255, 2013.

\bibitem{ding2011sensor}
Dan Ding, Rory~A Cooper, Paul~F Pasquina, and Lavinia Fici-Pasquina.
\newblock Sensor technology for smart homes.
\newblock {\em Maturitas}, 69(2):131--136, 2011.

\bibitem{hamad2020joint}
Rebeen~Ali Hamad, Longzhi Yang, Wai~Lok Woo, and Bo~Wei.
\newblock Joint learning of temporal models to handle imbalanced data for human
  activity recognition.
\newblock {\em Applied Sciences}, 10(15):5293, 2020.

\bibitem{ogbuabor2018human}
Godwin Ogbuabor and Robert La.
\newblock Human activity recognition for healthcare using smartphones.
\newblock In {\em Proceedings of the 2018 10th international conference on
  machine learning and computing}, pages 41--46, 2018.

\bibitem{chelli2019machine}
Ali Chelli and Matthias P{\"a}tzold.
\newblock A machine learning approach for fall detection and daily living
  activity recognition.
\newblock {\em IEEE Access}, 7:38670--38687, 2019.

\bibitem{saba2021suspicious}
Tanzila Saba, Amjad Rehman, Rabia Latif, Suliman~Mohamed Fati, Mudassar Raza,
  and Muhammad Sharif.
\newblock Suspicious activity recognition using proposed deep
  l4-branched-actionnet with entropy coded ant colony system optimization.
\newblock {\em IEEE Access}, 9:89181--89197, 2021.

\bibitem{sun2017sos}
Zehao Sun, Shaojie Tang, He~Huang, Zhenyu Zhu, Hansong Guo, Yu-e Sun, and
  Liusheng Huang.
\newblock Sos: Real-time and accurate physical assault detection using
  smartphone.
\newblock {\em Peer-to-Peer Networking and Applications}, 10(2):395--410, 2017.

\bibitem{antar2019challenges}
Anindya~Das Antar, Masud Ahmed, and Md~Atiqur~Rahman Ahad.
\newblock Challenges in sensor-based human activity recognition and a
  comparative analysis of benchmark datasets: a review.
\newblock In {\em 2019 Joint 8th International Conference on Informatics,
  Electronics \& Vision (ICIEV) and 2019 3rd International Conference on
  Imaging, Vision \& Pattern Recognition (icIVPR)}, pages 134--139. IEEE, 2019.

\bibitem{chen2021deep}
Kaixuan Chen, Dalin Zhang, Lina Yao, Bin Guo, Zhiwen Yu, and Yunhao Liu.
\newblock Deep learning for sensor-based human activity recognition: Overview,
  challenges, and opportunities.
\newblock {\em ACM Computing Surveys (CSUR)}, 54(4):1--40, 2021.

\bibitem{ni2015elderly}
Qin Ni, Ana Garc{\'\i}a~Hernando, and Iv{\'a}n de~la Cruz.
\newblock The elderly’s independent living in smart homes: A characterization
  of activities and sensing infrastructure survey to facilitate services
  development.
\newblock {\em Sensors}, 15(5):11312--11362, 2015.

\bibitem{dang2020sensor}
L~Minh Dang, Kyungbok Min, Hanxiang Wang, Md~Jalil Piran, Cheol~Hee Lee, and
  Hyeonjoon Moon.
\newblock Sensor-based and vision-based human activity recognition: A
  comprehensive survey.
\newblock {\em Pattern Recognition}, 108:107561, 2020.

\end{thebibliography}

\end{document}